\documentclass[amsmath,amssymb,11pt]{article}
\usepackage{jheppub2}
\pdfoutput=1
\usepackage{graphicx,epsfig}
\usepackage{float}
\usepackage{subfig}
\usepackage{subfloat}
\usepackage[normalem]{ulem}
\usepackage{diagbox}

\usepackage[utf8]{inputenc}
\usepackage{hyperref}
\usepackage{caption}
\usepackage{color}
\usepackage{overpic}
\usepackage[dvipsnames]{xcolor}
\usepackage{verbatim}
\usepackage{amsmath}

\newcommand*{\affmark}[1][*]{\textsuperscript{#1}}

\definecolor{pink1}{rgb}{0.858, 0.188, 0.478}

\newcommand{\beq}{\begin{equation}}

\newcommand{\eeq}{\end{equation}}

\title{Holographic thermodynamics of a charged AdS black holes with global monopole}

\author{He-Bin Zheng,\affmark[1]}
\emailAdd{zhenghb3060@163.com}
\author{Yun-Xian Chen,\affmark[1]}
\emailAdd{cyx17765580321@163.com}
\author{Jian Tang,\affmark[2*]}
\emailAdd{huzida@126.com}

\affiliation{\affmark[1] School of Physics and Astronomy, China West Normal University, Nanchong 637000, China}
\affiliation{\affmark[2] Department of physics and Electronic Science, Aba Teachers College, Wenchuan 623002, China}

\begin{abstract}
{By regarding the Newton constant $G_{N}$ and cosmological constant $\Lambda$ as variables, we in this paper study the thermodynamics and phase transition of Reissner-Nordstr$\ddot{o}$m anti-de Sitter (RN-AdS) black hole with global monopole in the framework of AdS/CFT correspondence.
It is found that there are interesting critical phenomena and phase behaviors in the (grand) canonical ensembles of fixed ($\tilde Q,{\cal V},C$), ($\tilde \Phi,{\cal V},C$) and ($\tilde Q,{\cal V},\mu$). When the other parameters are fixed, the free energy decreases with the global monopole increases. In ($\tilde Q,{\cal V},C$) ensemble, the range of unstable region decreases with the increase of global monopole. In ($\tilde \Phi,{\cal V},C$) ensemble, when $\tilde{\Phi} < \Phi_{c}$, the free energy appears two branches, the upper and lower branches correspond to low entropy and high entropy respectively. When ($\tilde Q,{\cal V},\mu$) is fixed, a new zero-order phase transition occurs in the high-entropy phase and the low-entropy phase at certain $\mu$-dependent temperatures. When $\mu$ increases to a certain value, this zero-order phase transition disappears. This certain value is negatively related to the magnitude of the global monopole. Finally, we find that $p-\cal V$ criticality does not appear with the change of global monopole. Therefore, it is important to note that the CFT states of charged black holes with global monopole do not correspond to Van der Waals fluids. Finally, we find that charged black holes with global monopoles can better reflect thermodynamic phase transitions and critical phenomena under the AdS/CFT correspondence. By adjusting the change of the global monopole, the thermodynamic phase transition will also change.}
\end{abstract}
\begin{document}

\maketitle

\section{Introduction}

Black holes are mysterious objects derived from Einstein's theory of general relativity. Because it will devour everything, light is no exception, so it is called ``black hole"\cite{Hawking:1975vcx}. For black holes, a large number of physicists and astronomers have devoted themselves to studying black holes from both experimental and theoretical aspects. Through the unremitting efforts of generations of researchers, the mystery of black holes has finally been uncovered. In theory, it is found that a black hole is a radiator, which can radiate energy outward, and is a thermodynamic system with temperature, volume, entropy, mass, etc. Until the 1970s, however, there was little reason for anyone to study the thermodynamics of black holes, until Hawking's area theorem changed this view\cite{Bardeen:1973gs,Bekenstein:1973ur}. Experimentally, the Laser Interferometer Gravitational-Wave Observatory(LIGO) has detected the gravitational wave signal generated by the merger of two massive black holes, providing effective evidence for the existence of black holes. Following this, the Event Horizon Telescope (EHT) reported images of supermassive black holes M87* and SgrA*\cite{AkiyamaL1,AkiyamaL2,AkiyamaL3,AkiyamaL4,AkiyamaL5,AkiyamaL6}, more directly proving the existence of black holes. Such a major breakthrough in both theory and experiment has made people more interested in scientific research.

It is well known that the laws of thermodynamics for ordinary matter always relate to the pressure-volume term, but the laws of thermodynamics for black holes do not. After continuous development, people regard the negative cosmological constant as the thermodynamic pressure P, and extend the thermodynamic space to the extended phase space. In physical terms, the cosmological constant can be regarded as a thermodynamic variable because Einstein's theory of gravity may be a low-energy efficient approximation to some more fundamental theory. On this basis, the physical constants in the efficient theory correspond to certain transverse parameters in the fundamental theory, thus allowing them to run in modular space. The development of extended phase space greatly enriches the thermodynamic phase transition and phase structure of black holes\cite{Kastor:2009wy,Dolan:2010ha,Dolan:2011xt,Cvetic:2010jb}. For example, polymorphism of mutual and phase transitions between black holes of different topologies, van der Waals phase transitions of AdS black holes\cite{Chamblin:1999tk,Chamblin:1999hg,Cvetic:1999ne,Kubiznak:2012wp}, and AdS black hole phase Spaces with more parameters. It is worth mentioning that when the cosmological constant is present, the mass of the black hole does not represent the internal energy of the system, but is interpreted as the enthalpy of the system. In our previous work, the study of thermodynamics in extended phase space has also made great progress.\cite{Zheng:2023ide,Mou:2024ecv,Mou:2023nrx,He:2024jdz,Li:2023ecv,Li:2022reh,Guo:2022yjc}.

Black hole chemistry has opened a new door for the study of black hole thermodynamics, and many new thermodynamic phase transitions have been generated, such as superfluid behavior\cite{Hennigar:2016xwd}, reentrant phase transitions\cite{Frassino:2014pha}, possible explanations of black hole microstructure and multi-critical phase transitions\cite{Wei:2019uqg,Tavakoli:2022kmo,Wu:2022plw}. One topic that has received a lot of attention in recent years is the holographic interpretation of black hole chemistry\cite{Kubiznak:2016qmn}. This interpretation holds that the thermodynamics of black holes in the body is equivalent to the thermodynamics of strongly coupled gauge theory at the boundary, given a large number of degrees of freedom in the limit of large N\cite{Kastor:2014dra,Dolan:2014cja,Zhang:2014uoa}. Changes in the volume cosmological constant will result in changes in the central charge and boundary volume of the CFT\cite{Karch:2015rpa}. In addition, both the CFT charge and the chemical potential (the conjugate quantity of the CFT charge) depend on the scale of the AdS length. Therefore, the first law of body thermodynamics cannot be projected directly onto dual CFT. In order to keep the central charge of the CFT able to vary independently, the researchers proposed a new method\cite{Visser:2021eqk,Cong:2021fnf,Cong:2021jgb}. This new approach is to introduce Newton's constant $G_{N}$ into the extended first law of thermodynamics as a variable parameter. In this new paradigm, no matter how the cosmological constant changes, the dual CFT does not change. Therefore, the first law of black hole body thermomechanics can be derived from the dual boundary field theory. Cong has explained the phase transition of RN-AdS using holographic thermodynamics through this new method\cite{Cong:2021fnf}. The degree of freedom of the dual field theory in the large N limit determines this transformation\cite{Kumar:2022fyq,Bai:2022vmx,Kumar:2022afq,Qu:2022nrt}.

Topological defects such as monopoles may arise during phase transitions\cite{Ahmed:2016ucs}. Monopole has a very high energy density. This intensity decreases with increasing distance. The relationship between intensity and distance is about $r^{-2}$\cite{Rhie:1990kc}. So at greater distances the total energy diverges. Having such a large energy intensity indicates that a global monopole can produce a strong gravitational field. Viewed from the global perspective of spacetime, the equatorial plane is a cone. The deficit Angle of the cone is $\Delta=8\pi^2 \zeta^2$\cite{Jusufik:2016glb}. It is easy to see that these monopoles exert gravity almost exclusively on relativistic matter\cite{Rhie:1990kc}. Barriola et.al discovery of the static solution of Einstein's equations with global monopoles\cite{Rhie:1990kc}. Yu pointed out that when the mass of the black hole devoring a monopole is large enough compared to the mass of the monopole, the presence of the global monopole will reduce the Hawking temperature and increase the area of the event horizon, and the relationship between entropy and the area of the event horizon will remain constant\cite{Yu:1994fy}. Many other authors have studied the physical properties of black holes with global monopoles\cite{Yu:2002st, Rahaman:2005sz, Pitelli:2009kd}. It is obvious that global monopoles play a very important role in the study of black holes. Therefore, on the basis of RN-AdS black holes, we will introduce the global monopole, an important parameter, to explore the relationship between it and holographic thermodynamics.

The purpose of this paper is to investigate the CFT phase behavior and critical phenomena. The line element uses RN-AdS black hole with global monopole. In section 2, extended thermodynamics and CFT thermodynamics are reviewed. In section 3, we discuss the free energy of regular ensembles of fixed ($\tilde Q,{\cal V},C$), ($\tilde \Phi,{\cal V},C$) and ($\tilde Q,{\cal V},\mu$), and study their phase behavior and critical phenomena. In section 4 we compare the $p-\cal V$ criticality of the CFT state with that of Van der Waals fluids. Heat capacity and thermal stability are also discussed in this section. Finally, in section 5, we discuss and summarize the full text. Through this study, we use the units $\hbar=\kappa_{\rm B}=c=1$.

\section{Holographic thermodynamics}
\label{sec:holographic}
 
 \subsection{Extended thermodynamics}
 
 In this paper, we consider the Einstein-Maxwell theory of four-dimensional spacetime\footnote[1]{A standard form of the action reads: $I=\frac{1}{16\pi G_{N}} \int \mathrm{d}^4 x \sqrt{-g}(\mathcal{R}-2\Lambda-G_{N}\mathcal{F}^2)$}
 \begin{equation}\label{eq:2.1}
 	I=\frac{1}{16\pi G_{N}} \int \mathrm{d}^4 x \sqrt{-g}(\mathcal{R}+K(\psi))\,,
 \end{equation}
 with
 \begin{equation}
     \psi=F_{\rm \mu\nu}F^{\mu\nu}\,,
 \end{equation}
 \begin{equation}
     F_{\rm \mu\nu}=\bigtriangledown_{\rm \mu}A_{\rm \nu}-\bigtriangledown_{\rm \nu}A_{\rm \mu}\,,
 \end{equation}
  the $G_{N}$ that appears in the above equation is Newton's constant, $\Lambda$ is a cosmological constant, $\mathcal{R}$ is a Ricci scalar, $A_{\mu}$ is the Maxwell field, $K(\psi)$ is the function of $\psi$, and $F_{\rm \mu \nu}$ denoting the electromagnetic field strength tensor. The convention (\ref{eq:2.1}) we use here will not change the physics\cite{Chamblin:1999tk}. The Lagrangian density of the charged AdS black hole with a global monopole is
  \begin{equation}
      \mathcal{L} = \mathcal{R} - 2\Lambda+\frac{1}{2}\partial_{\mu}\phi^a \partial^\mu \phi^{*a}-\frac{\gamma}{4}(\phi^a \phi^{*a}-\zeta_{0}^2)^2\,,
  \end{equation}
  where $\gamma$ is a constant, and $\zeta_{0}$ is the energy scale of symmetry breaking. The parameter $\phi^a$ represent a triplet of the scalar field
  \begin{equation}
      \phi^a = \zeta_{0}h(\tilde{r})\frac{\tilde{x}^a}{\tilde{r}}\,,
  \end{equation}
  where $\tilde{x}^a \tilde{x}^a = \tilde{r}^2$, $\tilde{x}^a$ is spatial Cartesian coordinates, $\tilde{r}$ is the radial distance, and $h(\tilde{r}) \rightarrow 1$ for $\tilde{r} \rightarrow \infty$. We consider the form of the asymptotic spherically symmetric metric for a RN-AdS black hole with a global monopole in four-dimensional spacetime\cite{Rhie:1990kc}
 \begin{equation} \label{eq:2.2}
 	\mathrm{d}s^2 = -f(\tilde{r})\mathrm{d}t^2+\frac{1}{f(\tilde{r})} \mathrm{d}\tilde{r}^2+\tilde{r}^2 \mathrm{d}\Omega_{2}^{2}\,,
 \end{equation} 
where $\mathrm{d}\Omega_{2}^{2}$ is the metric on the round unit 2-sphere, $f(\tilde{r})=1-8\pi \zeta_{0}^2-\frac{2\tilde{m}}{\tilde{r}}+\frac{\tilde{q}^2}{\tilde{r}^2}+\frac{\tilde{r}^2}{L^2}$, $\tilde{m}$ and $\tilde{q}$ correspond to mass and charge, respectively, and the black hole under consideration has no angular momentum. Introduce the following coordinate transformation
\begin{equation}
    \tilde{t} = (1-8\pi\zeta_{0}^2)^{-1/2}t\,,\qquad \tilde{r} = (1-8\pi\zeta_{0}^2)^{1/2}r\,,
\end{equation}
and new parameters
\begin{equation}
    m = (1-8\pi\zeta_{0}^2)^{-3/2}\tilde{m}\,,\qquad q = (1-8\pi\zeta_{0}^2)^{-1}\tilde{q}\,, \qquad \zeta^2 = 8\pi \zeta_{0}^2\,.
\end{equation}
Therefore, the metric in Eq.(\ref{eq:2.2}) can be rewritten as
\begin{equation}\label{dugui2}
    \mathrm{d}s^2 = -f(r)\mathrm{d}t^2 + \frac{1}{f(r)}\mathrm{d}r^2 + (1-\zeta^2)r^2(\mathrm{d}\theta^2+sin^2\theta \mathrm{d}\phi^2)\,,
\end{equation}
where
\begin{equation}
    f(r) = 1-\frac{2m}{r} + \frac{q^2}{r^2}+\frac{r^2}{L^2}\,.
\end{equation}
Take into account the monopole parameter $\zeta$ will result in a solid angle deficit in Eq.(\ref{dugui2}). The ADM mass and the electric charge of the black hole are modulated by the monopole parameter $\zeta$
 \begin{equation} \label{eq:2.3}
 	M=(1-\zeta^2) m \,,\qquad Q=(1-\zeta^2) q \,.
 \end{equation}
Fix the radius of the black hole at its outer horizon, if $f(r_{h})=0$, the mass parameter can be expressed in terms of other parameters
  \begin{equation} \label{eq:2.4}
 	m=\frac{q^2}{2r_{h}}+\frac{r_{h}}{2}+\frac{r{_h}{^3}}{2L^2}\,.
 \end{equation}
 Then the temperature, volume and electric potential at the outer event horizon are respectively
 \begin{equation}\label{eq:2.5}
  T=\frac{1}{4\pi r_{h}}\Bigg(1+\frac{3r_{h}^2}{L^2}-\frac{Q^2}{(1-\zeta^2)^2 r_{h}^2}\Bigg)\,, \qquad V = \frac{4\pi (1-\zeta^2) r_{h}^{3}}{3}\,, \qquad \Phi=\frac{q}{r_{h}}.
 \end{equation}
 The radius of curvature of AdS spacetime is $L$, which is related to the cosmological constant $\Lambda=-\frac{3}{L^2}\,.$ In extended phase space, the negative cosmological constant can be viewed as thermodynamic pressure $P=-\frac{\Lambda}{8\pi G_{N}}$\cite{Kastor:2009wy,Dolan:2010ha,Cvetic:2010jb,Dolan:2011xt,Kubiznak:2014zwa}.
In black hole thermodynamics, the black hole entropy $S$ and surface gravity $\kappa$ can be given by the following formula
\begin{equation}\label{eq:2.6}
 	S=\pi (1-\zeta^2)r_{h}^2 \,, \qquad \kappa=2\pi T \,.
\end{equation}

Taking cosmological constant $\Lambda$ and Newton constant $G_{N}$ into account, the first law of thermodynamics and Smarr relation of RN-AdS black holes in four-dimensional spacetime are given
  \begin{equation}\label{eq:2.7}
    \mathrm{d}M=\frac{\kappa}{8\pi G_{N}} \mathrm{d}A+\Phi \mathrm{d}Q+\frac{\Theta}{8 \pi G_{N}} \mathrm{d}\Lambda-(M-\Phi Q)\frac{\mathrm{d}G_{N}}{G_{N}},
 \end{equation}
 \begin{equation}\label{eq:2.8}
 	M=\frac{\kappa A}{4\pi G_{N}}+\Phi Q-\frac{\Theta \Lambda}{4 \pi G_{N}}.
 \end{equation}
 Here, $\Theta = - {\cal V}$ is the quantity conjugate to cosmological constant $\Lambda$, which can be defined as the background subtracted Killing volume\cite{Jacobson:2018ahi}
 \begin{equation}
     \Theta \equiv \int_{\Sigma_{bh}} |\xi|\mathrm{d}{\cal V} - \int_{\Sigma_{AdS}} |\xi|\mathrm{d}{\cal V}\,,
 \end{equation}
 where, $|\xi| = \sqrt{-\xi \cdot \xi}$ is the norm of the time translation Killing vector $\xi$. If we do not consider the change in Newton's constant. Combination Eq.(\Ref{eq:2.6}) and $P=-\frac{\Lambda}{8\pi G_{N}}\,,$ the extended first law of thermodynamics Eq.(\Ref{eq:2.7}) and the Smarr relation Eq.(\Ref{eq:2.8}) can be succinctly rewritten as
 \begin{equation} \label{eq:2.9}
   \mathrm{d}M=T\mathrm{d}S+\Phi \mathrm{d}Q+{\cal V}\mathrm{d}P,
 \end{equation}
 \begin{equation}\label{eq:2.10}
 	M=2TS+\Phi \mathrm{d}Q-2P{\cal V}.
 \end{equation}
However, when changes in Newtonian constants are added, the changes in cosmological and Newtonian constants in the first law of thermodynamics cannot be combined into a single term. Therefore, formula Eq.(\Ref{eq:2.7}) will be rewritten as
\begin{equation}\label{eq:2.11}
	\mathrm{d}M = \frac{\kappa}{8\pi }\mathrm{d}(\frac{A}{G_{N}})+\Phi \mathrm{d} Q+\frac{\Theta}{8\pi}\mathrm{d}(\frac{\Lambda}{G_{N}})-(M-\frac{\kappa A}{8\pi G_{N}}-\Phi Q-\frac{\Theta \Lambda}{8\pi G_{N}})\frac{\mathrm{d}G_{N}}{G_{N}}\,.
\end{equation}

 \subsection{CFT thermodynamics}
To embed the boundary center charge into the first law of thermodynamics Eq.(\Ref{eq:2.7}), it is necessary to use the holographic duality of the boundary central charge $C$, the bulk AdS scale $L$ and Newton's constant $G_{N}$, the form of the boundary central charge is obtained as follows\cite{Cong:2021fnf}
 \begin{equation}\label{eq:2.12}
 	C=\frac{\Omega_{2}L^2}{16\pi G_{N}}\,, 
 \end{equation}
 using the relationship between cosmological constant $\Lambda$, the bulk AdS scale $L$, Newton constant $G_{N}$ and pressure $P$, Newton constant $G_{N}$ is rewritten as
  \begin{equation}\label{eq:2.13}
         G_{N}= \frac{1}{4}\sqrt{\frac{3}{2\pi C P}}\,.
 \end{equation}
 Set the measurement of the CFT to\cite{Witten:1998qj,Ahmed:2023snm}
 \begin{equation}\label{eq:2.14}
 	\mathrm{d}s^2 = \omega^2(-\mathrm{d}t^2 + (1-\zeta^2)L^2 \mathrm{d}\Omega_{2}^{2})\,,
 \end{equation}
 $\omega$ is an arbitrary dimensionless conformal factor. The conformal factor is $\omega=R/L$ and $R$ is the radius of the boundary curvature. From this conformal factor, it is possible to establish a first law of holography containing a fixed Newtonian constant and a varying cosmological constant, which is completely dual to the first law of extended thermodynamics.
 
 The volume of the boundary space is proportional to $(1-\zeta^2)(\omega L)^2$, so write it in the form\cite{Cong:2021jgb}
 \begin{equation}\label{eq:2.15}
 	{\cal V} = (1-\zeta^2)\Omega_{2}R^2\,,
 \end{equation}
 the above equation gives the CFT volume, and naturally its conjugate CFT pressure should also be given. Thus, the $-p\mathrm{d}{\cal V}$ term appears. A holographic dictionary is used to correlate the entropy, mass, temperature, potential, and charge of black holes with their boundary counterparts\cite{Chamblin:1999tk,Karch:2015rpa,Visser:2021eqk}

 \begin{equation} \label{eq:2.16}
  S= \frac{A}{4G_N}, \qquad 	E = M \frac{L}{R},\qquad  T = \frac{\kappa}{2\pi} \frac{L}{R},\qquad \tilde \Phi =\frac{\Phi}{L}\frac{L}{R},  \qquad \tilde Q = Q L.
 \end{equation}
Due to the Casimir effect in four-dimensional spacetime, the vacuum energy of a CFT on a sphere is actually finite, and it has nothing to do with thermodynamic nature, so in the dictionary above, the vacuum state is considered to have no energy. The key step in matching the first law of volumes and boundaries is to replace $\Theta$ in the extended first law with the generalized Smarr formula and insert $\frac{\mathrm{d}\Lambda}{\Lambda}=-\frac{2\mathrm{d}L}{L}.$ After recombination, the extended first law can be expressed in terms of boundary thermodynamic quantities
 \begin{equation}
\begin{aligned} \label{eq:2.17}
	\mathrm{d} \left (  M \frac{L}{R} \right) &= \frac{\kappa   }{2\pi} \frac{L}{R} \mathrm{d} \left (  \frac{A}{4G_N}    \right  )   + \frac{\Phi}{R} \mathrm{d} (Q L )-          \frac{M}{2} \frac{L}{R} \frac{\mathrm{d} R^{2}}{R^{2}}  \\
	&\qquad + \left ( M \frac{L}{R} -   \frac{\kappa A}{8\pi G_N} \frac{L}{R}    - \frac{\Phi}{R} Q L  \right) \frac{\mathrm{d}\! \left ( L^{2}/G_N \right) }{ L^{2}/G_N}.
\end{aligned}
\end{equation}
By Eq.(\Ref{eq:2.12}), Eq.(\Ref{eq:2.15}) and Eq.(\Ref{eq:2.16}), it can be proved that the first law of holographic CFT extension is dual to the first law of extended thermodynamics
 \begin{equation}
 \label{eq:2.18}
 	\mathrm{d}E = T \mathrm{d}S + \tilde \Phi \mathrm{d} \tilde Q - p \mathrm{d} {\cal V} + \mu \mathrm{d}C.
 \end{equation}
By contrast Eq.(\Ref{eq:2.17}) and Eq.(\Ref{eq:2.18}), we can find CFT pressure and chemical potential
 \begin{equation}
 \label{eq:2.19}
 	p = \frac{1}{2} \frac{E}{ {\cal V}}\,, \qquad \qquad \mu  =\frac{1}{C}\left (  E - T S - \tilde \Phi \tilde Q \right).
 \end{equation}
In order to give the thermodynamic quantity of CFT more easily, two dimensionless parameters are introduced\cite{Dolan:2016jjc}
 \begin{equation}\label{eq:2.20}
 	x \equiv \frac{r_h}{L} \,,\qquad   y \equiv \frac{q}{L}.
 \end{equation}
Using the holographic dictionary Eq.(\Ref{eq:2.16}), the following CFT variables can be obtained:
\begin{equation}
\begin{aligned} \label{eq:2.21}
 	S = -4 C \pi x^2(-1+\zeta^2)\,,\qquad \tilde Q &=-4 C y (-1+\zeta^2) \,,\qquad \tilde \Phi =\frac{y}{R x} \,,\\
 	E = -\frac{2 C \Big(x^4+y^4+x^2\Big)(-1+\zeta^2)}{R x} \,,& \qquad T=\frac{(x^2+3x^4-y^2)(-1+\zeta^2)}{4\pi R x^3} \,,
 \end{aligned}
 \end{equation}
 similarly, the chemical potential $\mu$ is rewritten as
 \begin{equation}\label{eq:2.22}
 	\mu = \frac{(x^4+y^2-x^2)(-1+\zeta^2)}{R x},
 \end{equation}
 in the above equation, $R$ is the radius of boundary curvature and $C$ is the central charge of CFT.

 \section{CFT thermodynamic ensembles}
 \label{sec:ensembles}
 
In order to study the thermodynamic behavior of CFT states in an Einstein universe with global monopole, we selectively fix the charge $\tilde Q$, volume $\cal V$, central charge $C$, and chemical potential $\mu$ of a black hole to discuss its canonical ensemble.

We did find  phase transitions or  critical phenomena  
for the  remaining three ensembles at fixed $(\tilde Q, {\cal V}, C), (\tilde \Phi, {\cal V},C)$ and $(\tilde Q, {\cal V}, \mu)$, whose free energies we denote as $F\,,\, W$ and $ G\,,$ respectively,
   \begin{equation}
     \begin{aligned}  \label{eq:3.1}
    &\text{fixed} \quad  (\tilde Q, {\cal V}, C): \qquad &&F \equiv E - TS  = \tilde \Phi \tilde Q +\mu C \,,\\
      &\text{fixed} \quad  (\tilde \Phi, {\cal V},C)\,: \qquad &&W \equiv E- TS - \tilde \Phi \tilde Q =\mu C\,,\\
       &\text{fixed} \quad  (\tilde Q, {\cal V}, \mu)\,: \qquad &&G \equiv E - TS-     \mu C=\tilde \Phi \tilde Q\,. 
     \end{aligned}
 \end{equation}

 \subsection{Ensemble at fixed ($\tilde Q, {\cal V}, C$)}
 \label{sec:firstensemble}
 
 \begin{figure}
    \centering
    \includegraphics[scale=0.5]{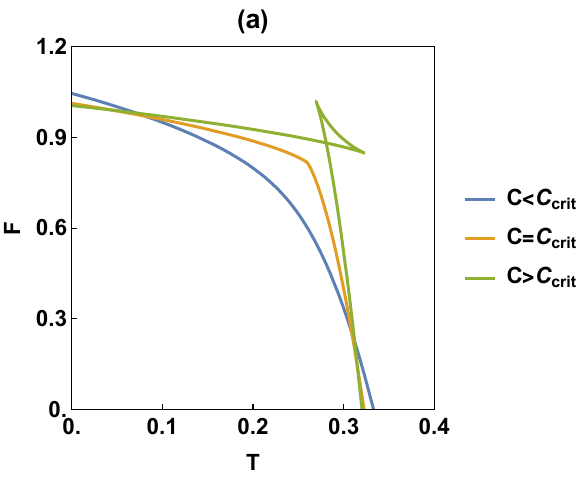}
    \includegraphics[scale=0.5]{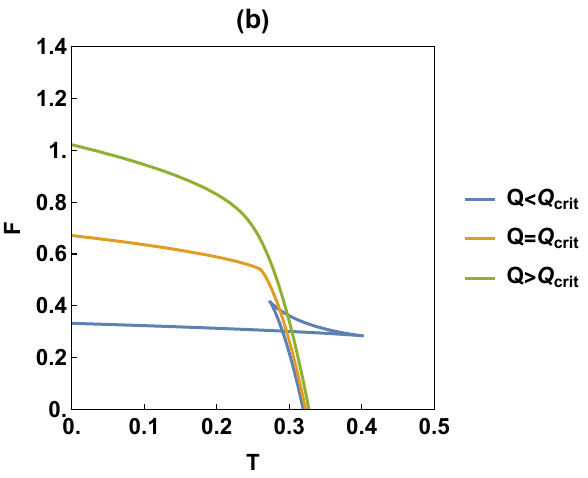} 
    \includegraphics[scale=0.5]{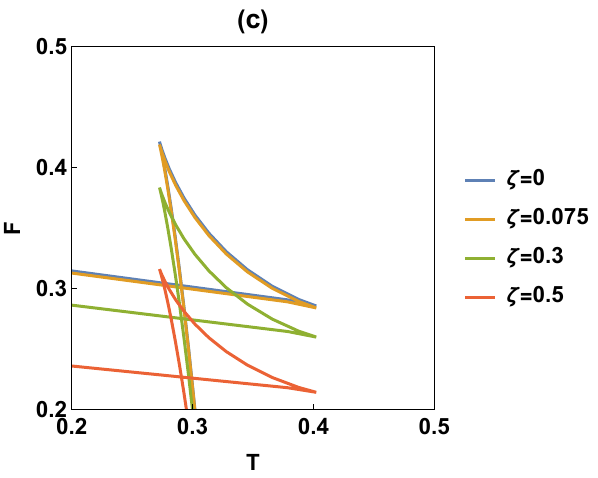}
    \caption{Free energy $F$ vs. temperature $T$ plot for the fixed $(\tilde Q,{\cal V},C)$ ensemble. Figure 1 (a), we plot different values of $C$ for fixed $\zeta$, the  parameters are $R=1$, $\tilde Q=1$, $\zeta=0.075$ and $C<C_{crit}, C=C_{crit}, C>C_{crit}$ (blue, orange, green). Figure 1 (b), we plot different values of $\tilde Q$ for fixed $\zeta$, the  parameters are $R=1$, $C=1$, $\zeta=0.075$ and $Q<Q_{crit}, Q=Q_{crit}, Q>Q_{crit}$ (blue, orange, green). Figure 1 (c), we plot different values of $\zeta$ for fixed $\tilde Q=1$, the parameters are $R=1$, $C<C_{crit}$ and $\zeta=0, 0.075, 0.3, 0.5$ (blue, orange, green, red).}
    \label{figure 1}
\end{figure}
 
In a canonical ensemble we fix the electric charge $\tilde Q$, the spatial volume ${\cal V}$ and the central charge $C$. The thermodynamic potential is  
\begin{align}
\label{eq:3.2}
 	F  \equiv E - T S 
 	 = \frac{C \Big(-x^2+x^4-3y^2\Big)(-1+\zeta^2)}{R x}.
\end{align}
Let us now examine how the free energy $F$ is a function of temperature $T$ at different fixed ($\tilde Q, {\cal V}, C$) values. For this purpose, it is practical to treat $F$ and $T$ as functions of ($\tilde Q, R, C, x$), and we note that the fixed radius $R$ is the same as according to the fixed volume $\cal V.$ Using relationship $\tilde Q = -4 C y (-1+\zeta^2),$ and then we get the free and the temperature, respectively,
 \begin{equation}\label{eq:3.3}
     F = \frac{C \Big(x^4-3(\frac{\tilde Q}{4C-4C \zeta^2})^2 - x^2\Big)(-1+\zeta^2)}{R x}\,,\qquad T = \frac{x^2+3x^4-(\frac{\tilde Q}{4C-4C \zeta^2})^2}{4 \pi R x^3}\,.
 \end{equation}
 Using free energy and temperature, we draw Figure \Ref{figure 1}, which shows the change of free energy under different global monopole for less than the critical value of charge and greater than the critical value of central charge, respectively, and the coexistence curve of free energy under different central charge values and charge values for given global monopole. In Figure \Ref{figure 1} (a), we keep the global monopole unchanged and change the value of the central charge. It can be found that when the central charge is greater than the critical value of the central charge, the black hole has a phase transition and a dovetail curve appears. When the central charge is less than or equal to its critical value, no phase transition occurs. In Figure \Ref{figure 1} (b), we keep the global monopole unchanged and change the value of charge. Different from Figure \Ref{figure 1} (a), phase transition occurs when the charge is less than the critical value of charge, and no phase transition occurs when the charge is greater than or equal to the critical value of charge. In Figure \Ref{figure 1} (c), the global monopole is changed by keeping the charge always lower than the critical value of the charge. The same conclusion can be obtained, with the increase of global monopole, the size of the coexistence zone does not change, and the temperature and free energy of the triple point decrease.

 \subsection{Ensemble at fixed ($\tilde\Phi, {\cal V}, C$)}
 
\begin{figure}[H]
 	\centering
 	\includegraphics[scale=0.5]{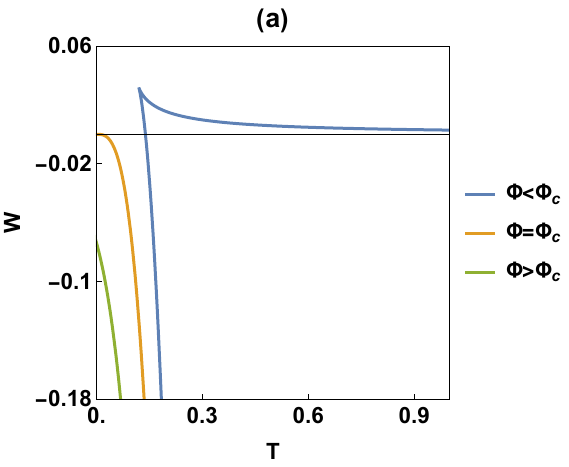} \hspace{0.1cm}
 	\includegraphics[scale=0.5]{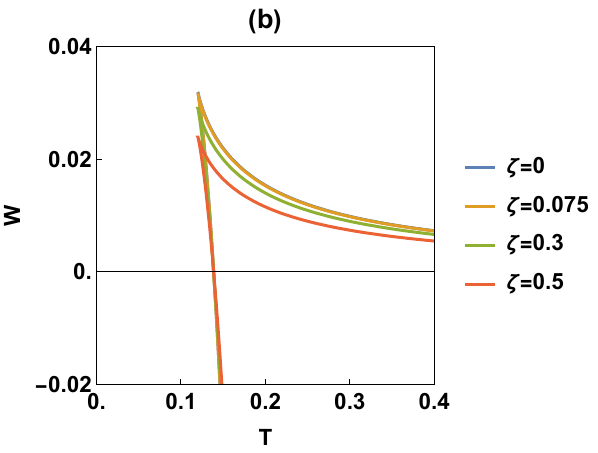}
 	\includegraphics[scale=0.5]{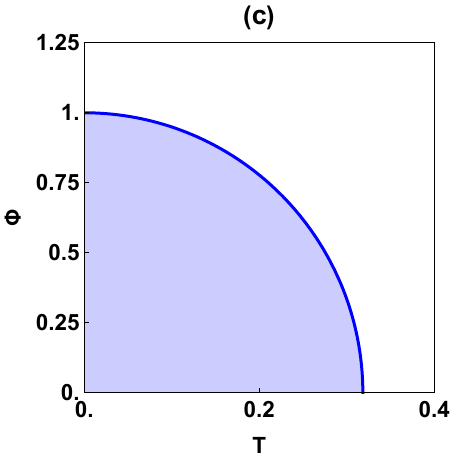}
 	\caption{Free energy $W$ vs. temperature $T$ plot and phase diagram for the fixed $(\tilde \Phi,{\cal V},C)$ ensemble.  \textbf{Left}: $W - T$ plot for the parameters  $R=1$, $C=1$, $\zeta=0.075$ and $\tilde\Phi<\Phi_c$ (blue), $\tilde\Phi=\Phi_c =  1$ (orange) and $\tilde\Phi>\Phi_c$ (green). \textbf{Middle}: $W - T$ plot for the parameters  $R=1$, $C=1$, $\tilde\Phi<\Phi_c$ and $\zeta=0, 0.075, 0.3, 0.5$(blue, orange, green, red).
 		\textbf{Right}: The  $\tilde \Phi-T$ phase diagram for $R=C=1$ with a coexistence curve  representing a line of (de)confinement phase transitions in the CFT.
 	} 
 	\label{figure 2}
 \end{figure}

In a grand canonical ensemble, we fix the potential $\tilde \Phi$, the volume $\mathcal{V}$ and the central charge $C$, and the thermodynamic potential is
\begin{equation} \label{eq:3.5}
 	W \equiv E - TS - \tilde \Phi \tilde Q = \frac{C \Big(x^4+y^2-x^2 (-1+2\zeta)\Big)}{R x}\,.
\end{equation}
Free energy $W$ and temperature $T$ are written as functions of $W(\tilde \Phi,C,\zeta,x,\cal V)$ and $T(\tilde \Phi,\zeta,x,\cal V)$, respectively.
 \begin{equation} \label{eq:3.6} 
 	W = \frac{C \Big(x^4+(R x \tilde \Phi)^2 - x^2\Big)(-1+\zeta^2)}{R x}\,,\qquad T = \frac{x^2+3x^4-(R x \tilde \Phi)^2}{4\pi R x^3}\,.
 \end{equation}
In Figure \Ref{figure 2} (a), leave the global monopole unchanged and change the potential. The $W-T$ diagram shows different behavior above and below a certain critical potential. For the case where the potential is greater than or equal to the critical value of the potential, the free energy is a smooth curve with no extreme value. When the potential is less than the critical value of the potential, the free energy appears a vertex, and there are two branches at the vertex, the upper and lower branches correspond to low entropy and high entropy respectively. When temperature is a function of $x$, there is a minimum at the vertex. In order to study the influence of global monopole on this phenomenon, we always keep the potential below the critical potential, and get Figure \Ref{figure 2} (b). With the increase of global monopole, the position of the vertex moves to the lower left, indicating that the critical position of low entropy and high entropy is constantly decreasing. The free energy of the lower branch is converted at $W = 0$, representing a first-order phase transition. When $W < 0$, the ``defined" state of high entropy predominates, while when $W < 0$, thermodynamically the ``limited" state is preferred. This (de-) confined phase transition is dual to the generalized Hawking Page phase transition between a large AdS black hole and an AdS spacetime with thermal radiation. Although the Hawk-Page transition was originally discovered in ADS-Schwarzschild black holes, it also exists in charged AdS black holes, where the transition depends on the value of the electric potential. This generalized Hawking Page transition exists both for Lifshitz black holes $1 \leqslant z \leqslant 2$(where $z$ is the dynamic Lifshitz index) and for any hyperscale violation parameter. Thus, there exists a complete line in the plane along which a first-order phase transition occurs between the restricted and defined phases.

By obtaining this first-order phase variation line analytically, set $W = 0$, and eliminating $x$ in favor of temperature, we can obtain the following expression for coexistence lines
\begin{equation} \label{eq:3.8}
	\tilde \Phi = \frac{\sqrt{1-\pi^2 R^2 T^2}}{R}\,,\qquad T_{c} = \frac{\sqrt{1 - R^2 \tilde \Phi^2}}{\pi R}\,.
\end{equation}
In Figure \Ref{figure 2} (c) we draw this first-order phase line on the $\tilde \Phi - T$ plane. When $T=0$, the phase transition occurs at $\tilde \Phi = \tilde \Phi_{c}$; When $\tilde \Phi = 0$, the phase transition is equivalent to the Hawking Page phase transition at $T = T_{c}$.

 \subsection{Ensemble at fixed ($\tilde Q, {\cal V}, \mu$)}
 \label{sec:thirdensemble}

Fixed charge $\tilde Q$, volume $\cal V$ and chemical potential $\mu$, thermodynamic potential is
\begin{equation} \label{eq:3.10}
	G \equiv E - T S - \mu C = -\frac{4 C y^2 (-1+\zeta^2)}{R x}.
\end{equation}
By combining the equations Eq.(\Ref{eq:2.21}) and Eq.(\Ref{eq:2.22}), the free energy $G$ and temperature $T$ are rewritten as $G(\tilde Q, {\cal V}, \mu, \zeta, x)$ and $T(\mu, \zeta, x, {\cal V})$.
\begin{equation}\label{eq:3.11}
	G = \frac{\tilde{Q}\sqrt{x(-1+\zeta^2)-x^3(-1+\zeta^2)+R \mu}}{R x\sqrt{\frac{-1+\zeta^2}{x}}}\,,
\end{equation}
\begin{equation}\label{eq:3.12}
	T = \frac{x}{\pi R} + \frac{\mu}{4\pi x^2 - 4\pi x^2 \zeta^2}\,.
\end{equation}

\begin{figure}[H]
	\centering 
	\includegraphics[scale=0.38]{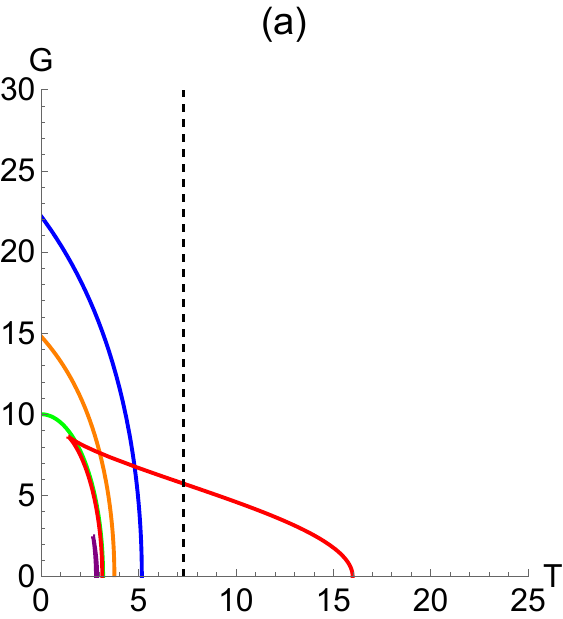} %\hspace{0.1cm}
	\includegraphics[scale=0.38]{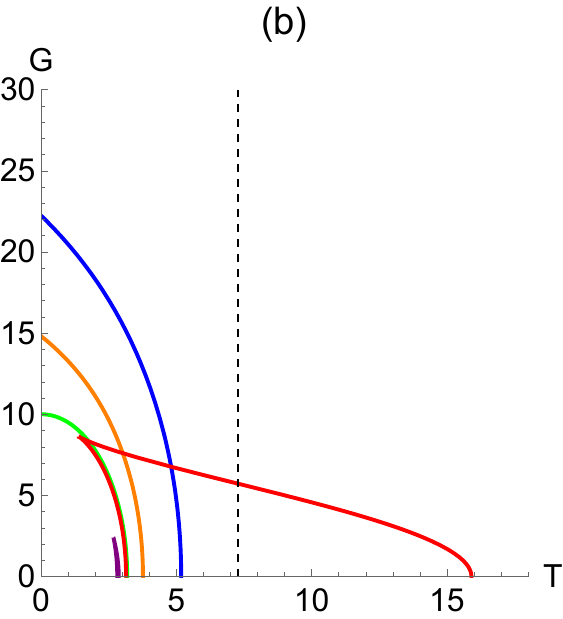}
	\includegraphics[scale=0.38]{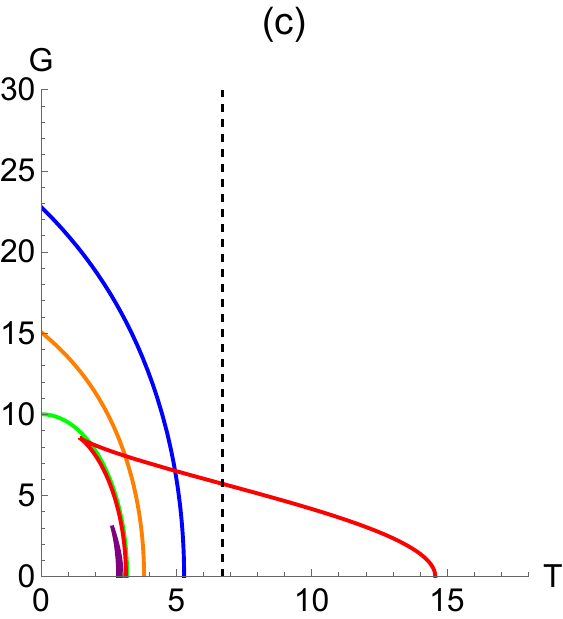}
	\includegraphics[scale=0.38]{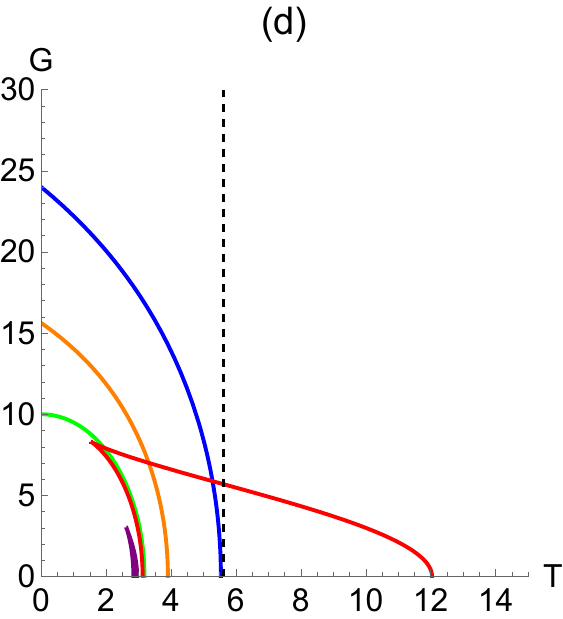}
	\includegraphics[scale=0.38]{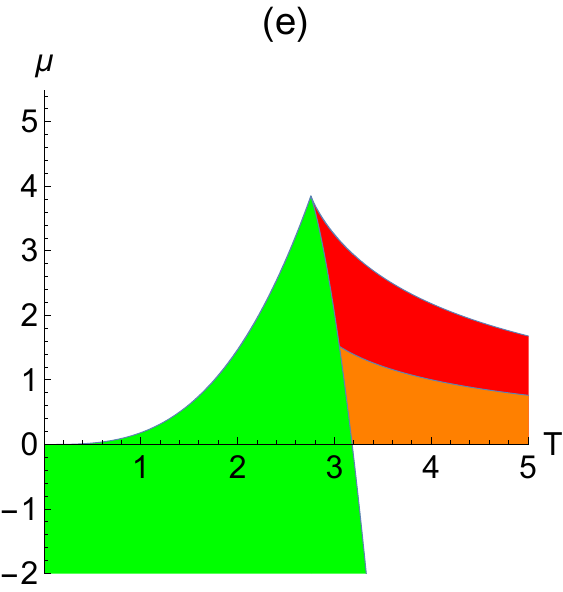}
	\includegraphics[scale=0.38]{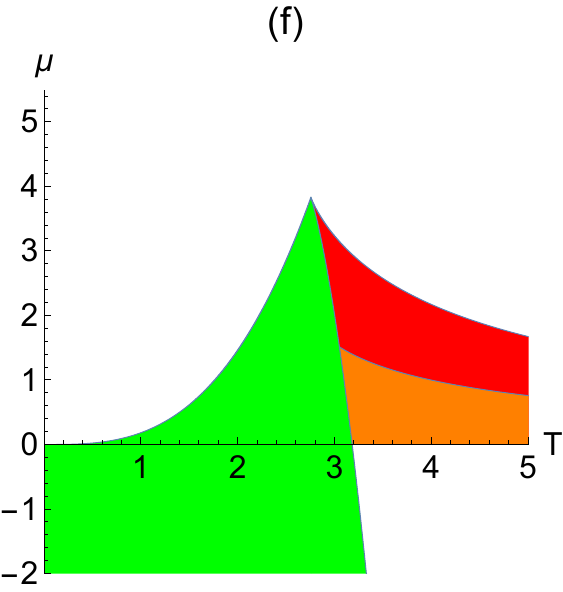}
	\includegraphics[scale=0.38]{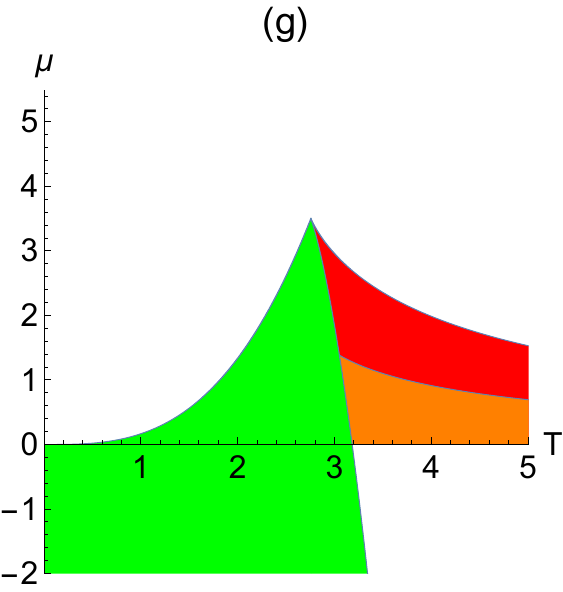}
	\includegraphics[scale=0.38]{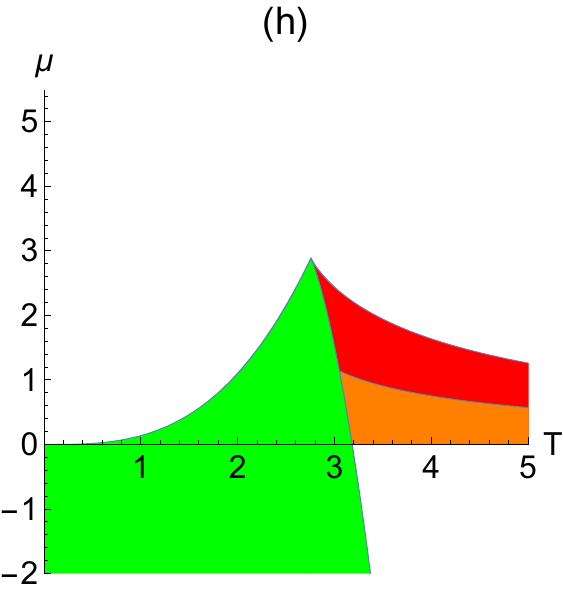}
	\caption{Free energy $G$ vs. temperature $T$ plot and phase diagram for the fixed $(\tilde Q,{\cal V},\mu)$ ensemble. In both figures we set $R=0.1$, $\tilde Q=1$. The $G-T$ plot for the values $\mu = -60,-10,0,1/10$(blue, orange, green, red) in the top row. The $\mu$ values of the purple curves in figure 3(a),3(b),3(c) and 3(d) are $\mu= 3.5,3.5,3,2.5$, respectively. The bottom row is the $\mu-T$ phase diagram corresponding to the top image. It shows demarcations between the different phases. In the top and bottom two rows of images, from left to right, $\zeta= 0, 0.075, 0.3, 0.5$.}
	\label{figure 3}
\end{figure}
\noindent
 In the top row, for $\mu\leq 0$ there is a monotonically stable phase. For $\mu>0$, the free energy curve splits into two branches. Both branches have a cusp in common at $T=T_{0}$. The temperature of the cusp can be computed by obtaining the stationary point of $T(\mu, \zeta, x, {\cal V})$
 \begin{equation}
 	(\frac{\partial T}{\partial x})_{\mu} = 0 \quad at \quad x_{0}=\frac{R^{1/3}\mu^{1/3}}{2^{1/3}(1-\zeta^2)^{1/3}}\,,\quad T_{0}=\frac{3 \mu^{1/3}}{2\pi R^{2/3}(2-2\zeta^2)^{1/3}}\,.
 \end{equation}
  \noindent
  And both branches are truncated by the line $G=0$ at $T_{1}$ and $T_{2}$ ($T_{1} \leq T_{2}$). $T_{1}$ and $T_{2}$ correspond to the two positive roots of the function $f(x) = \frac{\sqrt{x(-1+\zeta^2)-x^3(-1+\zeta^2)+R \mu\mu}}{R x\sqrt{\frac{-1+\zeta^2}{x}}}$. The high entropy state corresponds to the lower branch, and the low entropy state corresponds to the upper branch. For $T_{0}<T<T_{1}$ the high entropy phase plays a dominant role. At $T = T_{1}$, when the temperature rises, the high-entropy phase disappears and the CFT undergoes a zero-order phase transition into a low-entropy phase. There's an intermediate temperature $T_{int}$ (where the black dashed line intersects the red curve). At $T=T_{2}$, the low-entropy has positive heat capacity. From $T_{2}$ to $T_{1}$, the heat capacity becomes negative at the intermediate temperature. As the global monopole increases, $T_{1}$, $T_{2}$ and the intermediate temperature $T_{int}$ all decrease. And the range of high and low entropy states is also decreasing. In the bottom row, the green area represents the high entropy phase and the heat capacity is positive. Both the red and orange parts represent low-entropy phases. However, the red part has a positive heat capacity and is a stable phase. The orange part has a negative heat capacity and is an unstable phase. White regions indicate that no solution exists. With the increase of global monopole, the regions of high entropy phase, stable low entropy phase and unstable low entropy phase will decrease. When the temperature range is fixed and the global monopole is increased, the high entropy phase gradually takes the dominant position.

 \section{CFT criticality and thermodynamic stability}
\label{sec:critistabl}

\subsection{Critical points and comparison with Van der Waals fluid}
   
\begin{figure}
 	\centering
 	\includegraphics[scale=0.55]{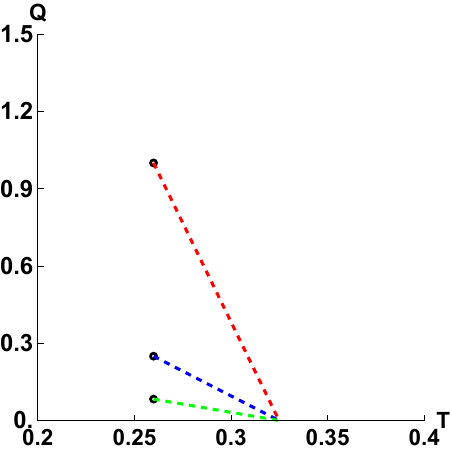} \hspace{1 cm}
 	\includegraphics[scale=0.55]{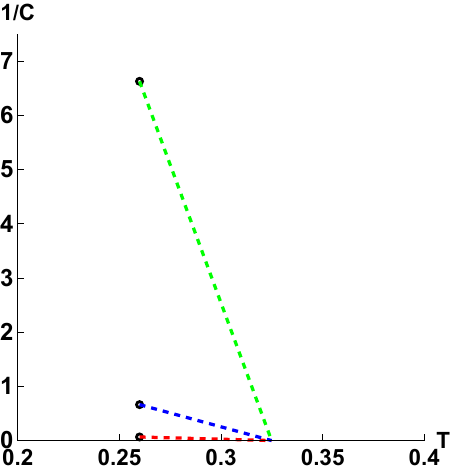}
 	\caption{Coexistence lines for the fixed $(\tilde Q,{\cal V},C)$ ensemble.  \textbf{Left}: Low-entropy and high-entropy coexistence curve for CFT thermal states on $\tilde Q-T$ phase diagram. The parameters used here are $R=1$, $C=1/10$ (red), $C=1$ (blue) and $C=10$ (green). For each value of $C$, the coexistence line represents a line of first-order phase transitions between low-entropy states (to the left of the line) and  high-entropy states (to the right), and the line ends at a critical point where a second-order phase transition occurs at $\tilde Q=\tilde Q_{crit}$ and $T=T_{crit}$.  \textbf{Right}: A similar coexistence curve of the low- and high-entropy CFT states exists in the $1/C-T$ phase diagram, which we depicted here for  $R =1$ and  $\tilde Q=1/2$ (red), $\tilde Q=2$ (blue) and $\tilde Q=6$ (green).} 
 	\label{figure 4}
 \end{figure}

 First calculate the value of the relevant thermodynamic quantity at the critical point. Temperature has an inflection point at the critical point as a function of $x$,i.e.
 \begin{equation} \label{eq:4.1}
 	\left (\frac{ \partial T}{\partial x} \right)_y= 0  =	\left ( \frac{\partial^2 T }{\partial x^2} \right)_y\, \qquad \text{at} \quad  x=x_{crit}  \quad \text{and} \quad y=y_{crit}.
 \end{equation}
 \begin{equation}
\label{eq:4.2}
 	x_{crit}  =	\frac{1}{\sqrt{6}},\qquad   y_{crit}  = \frac{1}{\sqrt{6}} x_{crit}.
 \end{equation}
From this, you can write the critical points of entropy $S_{crit}$, temperature $T_{crit}$, chemical potential $\mu_{crit}$, electric potential $\tilde \Phi_{crit}$, and pressure $p_{crit}$
\begin{equation} \label{eq:4.3}
	\begin{aligned}
	S_{crit} = \frac{2}{3} C \pi (1-\zeta^2) &\,, \qquad T_{crit} = \frac{\sqrt{\frac{2}{3}}}{\pi R} \,, \qquad \mu_{crit} = \frac{\sqrt{\frac{2}{3}}(1-\zeta^2)}{3 R}\,.\\& \tilde \Phi_{crit} = \frac{1}{\sqrt{6}R}\,, \qquad p_{crit} = \frac{C}{3\sqrt{6}\pi R^3}\,.
	\end{aligned}
\end{equation}
In Table \ref{table 1} and Table \ref{table 2}, we selected several different $\zeta$ values and obtained the corresponding slope values. In Figure \ref{figure 4}, we take $\zeta=0.075$ as an example to show Table \ref{table 1} and Table \ref{table 2} graphically. First of all, we can easily see that all slopes are negative. Figure \ref{figure 4} shows the coexistence lines of the low-entropy and high-entropy phases of the CFT on the $\tilde Q-T$ and $1/C-T$ phase diagrams. The coexistence line separates the two phases on the plane, and when the CFT crosses the coexistence line, it undergoes a first-order phase transition. On these two phase diagrams, the low-entropy phase is to the left of the coexistence line (for any given $C$ and $R$ values), while the high-entropy phase is to the right of the coexistence line. The critical point is represented on the graph by an open circle. Above the critical point, the CFT does not show significant phase. In the image of $\tilde Q-T$, the fixed charge $\tilde Q$ does not change, and the slope increases as $\zeta$ increases. The fixed charge $\zeta$ does not change, and as $\tilde Q$ increases, the slope increases. In the image of $1/C-T$, the fixed the central charge $C$ does not change, and the slope increases as $\zeta$ increases. The fixed charge $\zeta$ does not change, and as $1/C$ increases, the slope decreases.
 \begin{table}[h]
	\centering
	\begin{tabular}{|c|c|c|c|c|}
		\hline
		\diagbox{$\tilde Q$}{$\zeta$} & 0 & 0.075 & 0.3 & 0.5 \\
		\hline
		0.5 & -20.5208 & -20.4054 & -18.6739 & -15.3906 \\
		2 & -5.1302 & -5.10134 & -4.66848 & -3.84765 \\
		6 & -1.71007 & -1.70045 & -1.55616 & -1.28255 \\
		\hline
	\end{tabular}
	\caption{Slope of CFT hot low-entropy and high-entropy coexistence curves on $\tilde{Q}-T$ phase diagrams}
	\label{table 1}
\end{table}

\begin{table}[h]
	\centering
	\begin{tabular}{|c|c|c|c|c|}
		\hline
		\diagbox{$1/C$}{$\zeta$} & 0 & 0.075 & 0.95 & 1.95 \\
		\hline
		0.1 & -1.02604 & -1.02027 & -0.933696 & -0.76953 \\
		1 & -10.2604 & -10.2027 & -9.33696 & -7.6953 \\
		10 & -102.604 & -102.027 & -93.3696 & -76.953 \\
		\hline
	\end{tabular}
	\caption{Slope of CFT hot low-entropy and high-entropy coexistence curves on $1/C-T$ phase diagrams}
	\label{table 2}
\end{table}
 
\begin{figure}[H]
 	\centering
 	\includegraphics[scale=0.55]{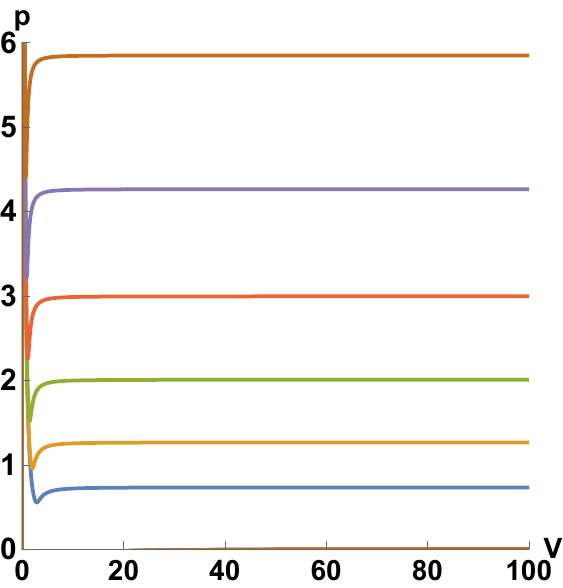} %\hspace{1 cm} 
 	\caption{No $p-{\cal V}$ criticality for holographic CFTs. The phase diagram of fixed $\zeta = 0.075$, gradually increasing the temperature($T=0.5,0.6,0.7,0.8,0.9,1$). The parameters used here are $\tilde Q =C= 1$.}
 	\label{figure 5}
 \end{figure}

 Figure \ref{figure 5} shows the phase diagram of CFT thermodynamics. It is important to note that there is no $p-{\cal V}$ criticality in CFT thermodynamics. This can be easily seen in the image on the left of Figure \ref{figure 5}. No matter how the temperature changed, a similar picture appeared(no critical turning point). 

%%%%%%%%%%%%%%%%%%%%%%%%%%%%%%%%%%%%%%%%%%%%%%%%%%%%%%%%%%%%%%%%%%%%%%%%%%%%%%%%%%%%%%%
%%%%%%%%%%%%%%%%%%%%%%%%%%%%%%%%%%%%%%%%%%%%%%%%%%%%%%%%%%%%%%%%%%%%%%%%%%%%%%%%%%%%%%%
\subsection{Critical exponents for fixed ($\tilde Q, {\cal V}, C$) ensemble}

In this section, we discuss the criticality in the $C-\mu$ plane. First, the central charge and the chemical formula are written as follows
\begin{equation}
    C = \frac{\tilde Q}{4\sqrt{x^2(1-4\pi R T x+3x^2)}(1-\zeta^2)}\,,
\end{equation}
\begin{equation}
    \mu = \frac{4(\pi R T -x)x^2(1-\zeta^2)}{R}\,.
\end{equation}
\begin{figure}[h!]
 	\centering
 	\includegraphics[scale=0.55]{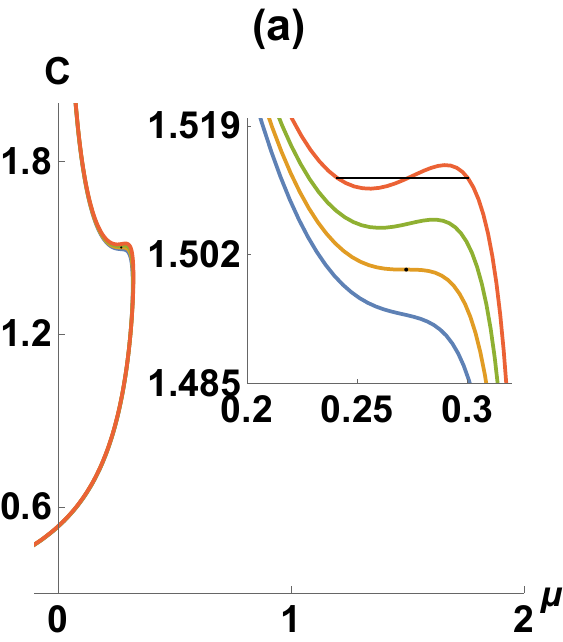} \hspace{0.5 cm}
 	\includegraphics[scale=0.55]{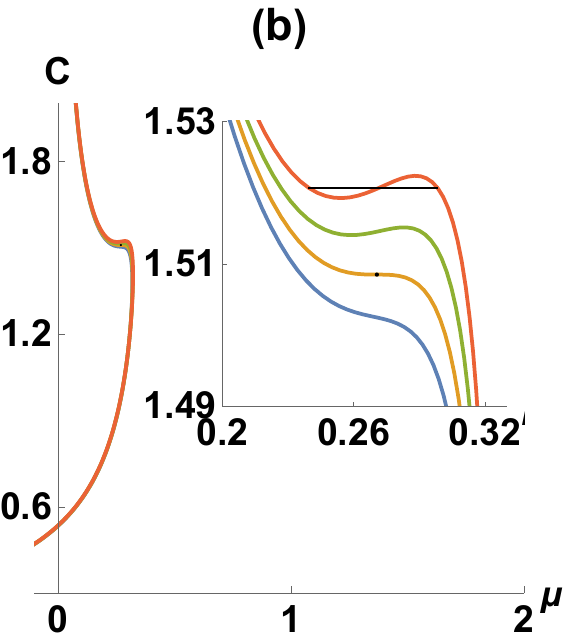} \\
        \includegraphics[scale=0.55]{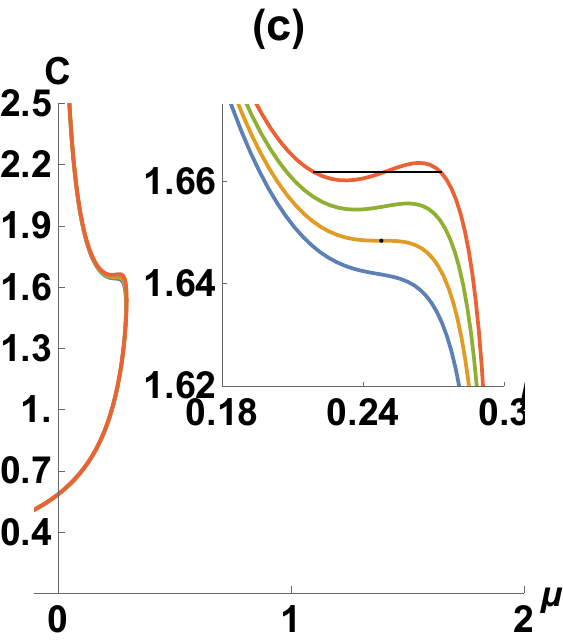} \hspace{0.5 cm}
        \includegraphics[scale=0.55]{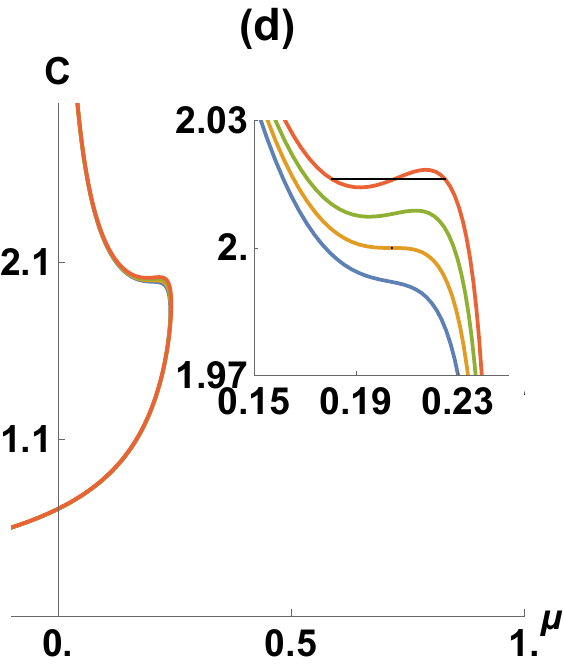}
 	\caption{Isotherms in the $C-\mu$ plane. The parameters used here are $\tilde Q = 1$, $R = 1$, $T=T_{crit}=\frac{1}{\pi}\sqrt{\frac{2}{3}}$(orange), $T=0.999T_{crit}$(blue), $T=1.001T_{crit}$(green) and $T=1.002T_{crit}$(red). For $T>T_{crit}$, the curve displays a ``wriggle" similar to the $P-\cal V$ phase diagram in the Van der Waals liquid-gas phase transition.}
 	\label{figure 44}
 \end{figure}
In Figure \ref{figure 44}, we plot $C-\mu$ images at different temperatures and different global monopoles. In Figure \ref{figure 44}, images from (a) to (d) correspond to $\zeta= 0, 0.075,
 0.3, 0.5$, respectively. It is not difficult to see that each curve will appear a chemical potential maximum position. At position $x = \frac{2\pi R T}{3}$, maximum chemical potential $\mu_{max}=\frac{16}{27}\pi^3 R^2 T^3 (1-\zeta^2)$ can be obtained. We used a black dot to represent the inflection point of the curve. Above the critical temperature, a phase transition occurs in the curve. To see the phase transition more clearly, we use Maxwell's law of equal area to divide this curved site into upper and lower parts using a horizontal line. The first phase transition of fixed (${\tilde Q}, {\cal V}, C$) ensemble occurs on this transverse line. We find that the central charge increases with the increase of the global monopole. The high and low entropy phases of the first order phase transition can be controlled by changing the global monopole.

In order to study the critical exponents in the $C-\mu$ plane, the variables are defined
\begin{align}
    t \equiv \frac{T-T_{crit}}{T_{crit}}\,,\qquad \chi \equiv \frac{C}{C_{crit}}\,, \qquad \psi \equiv \frac{\mu-\mu_{crit}}{\mu_{crit}}\,.
\end{align}
With these new parameters, we can describe the behavior of some physical quantities, for example, heat capacity ${\cal C}_{\tilde Q, {\cal V}, \mu}$, the order parameter $\eta \equiv \mu_h - \mu_l$ and the isothermal compressibility-like quantity $\kappa_T$ at the vicinity of the critical point
\begin{equation}\label{4.7}
    {\cal C}_{\tilde Q, {\cal V}, \mu} \equiv T \left ( \frac{\partial S}{\partial T}\right)_{\tilde Q, {\cal V}, \mu} \sim \left| t \right|^{-\alpha}\,,
\end{equation}
\begin{equation}
    \eta \equiv \mu_h - \mu_l \sim \left| t \right|^{\beta}\,,
\end{equation}
\begin{equation}
    \kappa_T \equiv -\frac{1}{\mu}\left(\frac{\partial \mu}{\partial C}\right)_{T, \tilde Q, \cal{V}} \sim \left| t \right|^{-\gamma}\,.
\end{equation}
Near the critical point of temperature $T_{crit}$, the central charge behaves as
\begin{equation}
    \left|C-C_{crit}\right| \sim \left| \mu - \mu_{crit} \right|^{\delta}\,,
\end{equation}
the above equation introduces a new critical exponent $\delta$. To make it easier to find $\alpha$, we rewrite $S$ and $T$ in the following form
\begin{equation}
   S = \frac{\pi {\tilde Q} x^2 \sqrt{\frac{-1+\zeta^2}{x}}}{\sqrt{-x(-1+x^2)(-1+\zeta^2)+R \mu}}\,,
\end{equation}
\begin{equation}
   T = \frac{x}{\pi R}+\frac{\mu}{4\pi x^2-4\pi x^2 \zeta^2}\,.
\end{equation}
Eq.(\ref{eq:2.21}) and Eq.(\ref{eq:2.22}) are introduced, and Eq.(\ref{4.7}) are rewritten as follows
\begin{equation}
    \begin{aligned}
    {\cal C}_{\tilde Q,{\cal V},\mu} &= \frac{\pi \tilde{Q}x^2\sqrt{\frac{-1+\zeta^2}{x}}\Big(4x^3(-1+\zeta^2)-R \mu\Big)\Big(2x(-1+\zeta^2)+3R \mu \Big)}{4\Big(2x^3 (-1+\zeta^2)+R \mu \Big)\Big(-x(-1+x^2)(-1+\zeta^2)+R\mu\Big)^{3/2}} 
    \\ &\xrightarrow[\substack{x\to x_{crit} \\ \mu\to \mu_{crit}}]{}\,\,\,\,\,\, 0 \quad \implies \quad \alpha = 0\,,
    \end{aligned} 
\end{equation}
the heat capacity converges at the critical position, so $\alpha = 0$. We represent Eq.(\ref{4.7}) by $\chi = \chi(t, \psi)$ and unfold around the critical point $t=\psi=0$. This gives us
\begin{equation}
   \chi = 1+4t+12t \psi-2\psi^3+O(t \psi^2,\psi^4) \equiv 1+\overline{A}t+\overline{B}t \psi+\overline{C}\psi^3+O(t \psi^2,\psi^4)\,,
\end{equation}
$\overline{A}$, $\overline{B}$ and $\overline{C}$ in the above equation are the number of generations. For first-order phase transitions near critical points in the $C-\mu$ plane, we have Maxwell's equal area law
\begin{equation}
    \int_{\psi_l}^{\psi_h} \psi \rm{d} C=0\,,
\end{equation}
where
\begin{equation}
    \begin{aligned}
      \psi_h \equiv \frac{\mu_h}{\mu_c}-1\,, \qquad \psi_l \equiv \frac{\mu_l}{\mu_c}-1\,.
    \end{aligned}
\end{equation}
On the isotherm, we have
\begin{equation}
    \rm{d}C = C_{crit}\left(\overline{B}t+3\overline{C}\psi^2\right)\rm{d}\psi\,.
\end{equation}
In addition, we also have a constant $\chi$ in the first-order phase transition near the critical point
\begin{equation}
    \chi = 1+\overline{A}t+\overline{B}t \psi_h+\overline{C}\psi{_h^3}=1+\overline{A}t+\overline{B}t \psi_l+\overline{C}\psi{_l^3}\,,
\end{equation}
\begin{equation}
  \int_{\psi_l}^{\psi_h} \psi \left(\overline{B}t+3\overline{C}\psi^2\right) \rm{d} \psi=0\,.
\end{equation}
There is a unique nontrivial solution
\begin{equation}
   \psi_h = -\psi_l = \sqrt{\frac{-\overline{B}t}{\overline{C}}}\,.
\end{equation}
Can give immediately
\begin{equation}
    \eta = \mu_{crit}\left(\psi_h-\psi_l\right) = 2\mu_{crit}\sqrt{\frac{-\overline{B}t}{\overline{C}}}\,.
\end{equation}
So we know the exponent $\beta = 1/2$. To calculate $\gamma$, by taking the derivative and we can get
\begin{equation}
   \left(\frac{\partial \mu}{\partial C}\right)_{T, {\tilde Q}, \cal V} = \frac{\mu_{crit}}{y_{crit}\overline{B}}\frac{1}{t}+O(\psi)\,.
\end{equation}
can be further obtained
\begin{equation}
   \kappa_T \sim \frac{1}{y_{crit}\overline{B}t}\,,
\end{equation}
So we know the exponent $\gamma = 1$. Again, set $t=0$ and we get
\begin{equation}
    \chi = 1 + \overline{C}\psi^3\,.
\end{equation}
That's how we know $\delta = 3$. As we can see, the critical index, the index of the scaling law that describes the behavior of CFT states near critical points, is the same as the index of charged AdS black holes previously discovered. This proves that the critical index of CFT criticality belongs to the class of universality predicted by the mean field theory.

  \subsection{Heat capacities and thermal stability}

In order to study the thermodynamic phase transition and critical behavior of CFT, only the thermodynamic stability of three main ensembles is considered in this paper. The heat capacities of different ensembles are defined as follows
 \begin{align}
 	{\cal C}_{\tilde Q,{\cal V},C} &\equiv T \left ( \frac{\partial S}{\partial T}\right)_{\tilde Q, {\cal V}, C} =  \left ( \frac{\partial E }{\partial T }\right)_{\tilde Q, {\cal V}, C}\,,\\
  {\cal C}_{\tilde\Phi,{\cal V},C}  &\equiv T \left ( \frac{\partial S}{\partial T}\right)_{\tilde\Phi, {\cal V}, C} =   \left( \frac{\partial (E- \tilde\Phi \tilde Q) }{ \partial T   }  \right)_{\tilde\Phi, {\cal V}, C}\,,\\
 	{\cal C}_{\tilde Q,{\cal V},\mu}  &\equiv T \left ( \frac{\partial S}{\partial T}\right)_{\tilde Q, {\cal V}, \mu} = \left ( \frac{\partial (E- \mu C) }{ \partial T  }  \right)_{\tilde Q, {\cal V}, \mu}\,.
\end{align} 
Taking the derivative of temperature and entropy by the chain rule gives an expression for the heat capacity as a function of ($\tilde Q,{\cal V},C,\zeta,x$), ($\tilde \Phi,{\cal V},C,\zeta,x$), and ($\tilde Q,{\cal V},\mu,\zeta,x$), respectively.
 \begin{align}
 	{\cal C}_{\tilde Q,{\cal V},C} &= \frac{8 C \pi x^2(-1+\zeta^2)\Big(-\tilde{Q}^2+16C^2 x^2(1+3x^2)(-1+\zeta^2)^2\Big)}{3\tilde{Q}^2+16C^2 x^2(-1+3 x^2)(-1+\zeta^2)^2}\,, \\
   	{\cal C}_{\tilde\Phi,{\cal V},C} &= \frac{8 C \pi x^2(-1+\zeta^2)(-1-3x^2 +R^2 \tilde{\Phi}^2)}{-1+3x^2 +R^2 \tilde{\Phi}^2}\,,\\
  	{\cal C}_{\tilde Q,{\cal V},\mu} &= \frac{\pi \tilde{Q}x^2\sqrt{\frac{-1+\zeta^2}{x}}\Big(4x^3(-1+\zeta^2)-R \mu\Big)\Big(2x(-1+\zeta^2)+3R \mu \Big)}{4\Big(2x^3 (-1+\zeta^2)+R \mu \Big)\Big(-x(-1+x^2)(-1+\zeta^2)+R\mu\Big)^{3/2}}\,.
   \end{align}

\begin{figure}[H]
 	\centering
 	\includegraphics[scale=0.55]{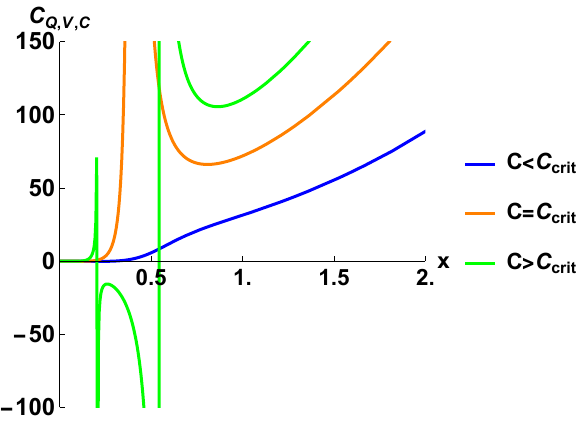}\hspace{0.5 cm}
 	\includegraphics[scale=0.55]{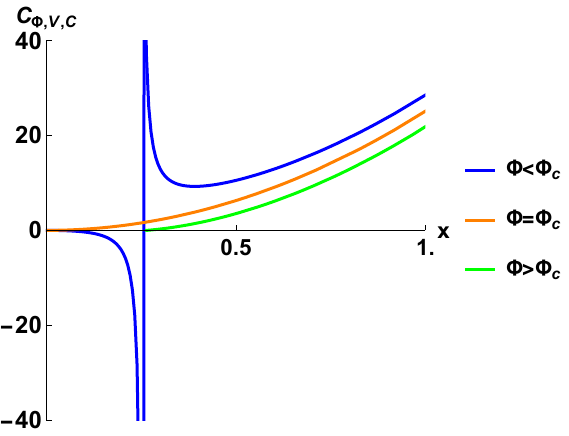}
 	\caption{Heat capacities, for fixed $(\tilde Q,{\cal V},C)$ and $(\tilde\Phi ,{\cal V},C)$ ensembles. Here we use the same parameters as in Figures \ref{figure 1}, \ref{figure 2}, and \ref{figure 3} for easy observation and comparison.}
 	\label{figure 6}
 \end{figure}

 We plot these heat capacities in Figure \ref{figure 6}. Corresponding to fixed $(\tilde Q,{\cal V},C)$ and $(\tilde\Phi ,{\cal V},C)$ ensembles from left to right. In the fixed $(\tilde Q,{\cal V},C)$ ensemble, $C > C_{crit}$ corresponds to the green curve. The curves correspond to low, intermediate and high entropy of CFT states. The heat capacities of both the low and high entropy states are positive, and both states are stable. However, the intermediate entropy state is not stable, and its heat capacity is negative. $C > C_{crit}$ corresponds to the orange curve. In this case, the heat capacity is always positive. At the critical point $x_{crit} = \frac{1}{\sqrt{6}}$, the heat capacity goes to infinity. $C < C_{crit}$ corresponds to the blue curve. At this time, the heat capacity of the CFT state is always positive, and there is no noteworthy thermodynamic phase transition. The critical phenomena corresponding to this row of images are exactly the same as in Figure \ref{figure 1}.
 
 The images in the right of Figure \ref{figure 6} correspond to the $(\tilde\Phi ,{\cal V},C)$ ensemble. When $\tilde {\Phi} < \Phi_{c}$, the curve appears two branches, corresponding to the blue curve in the figure. When $x$ is small, the heat capacity of the CFT state is negative, corresponding to the low entropy branch in Figure \ref{figure 2}, and the heat state is unstable at this time. When $x$ is large, the heat capacity is positive, corresponding to the high entropy branch in Figure \ref{figure 2}, which has a stable hot state. The critical point between the low entropy state and the high entropy state is $x_{cusp} = \frac{\sqrt{1-R^2 \Phi^2}}{\sqrt{3}}$. The heat capacity diverges around $x_{cusp}$. When $\tilde {\Phi} \geq \Phi_{c}$ (orange, green), the heat capacity is always positive.
 
\begin{figure}[H]
 	\centering
 	\includegraphics[scale=0.55]{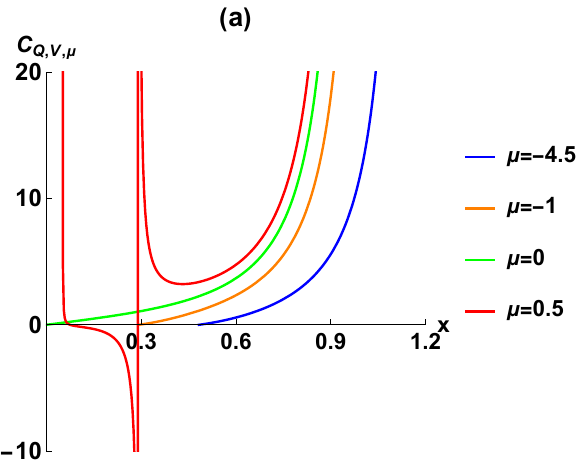}\hspace{0.5 cm}
 	\includegraphics[scale=0.55]{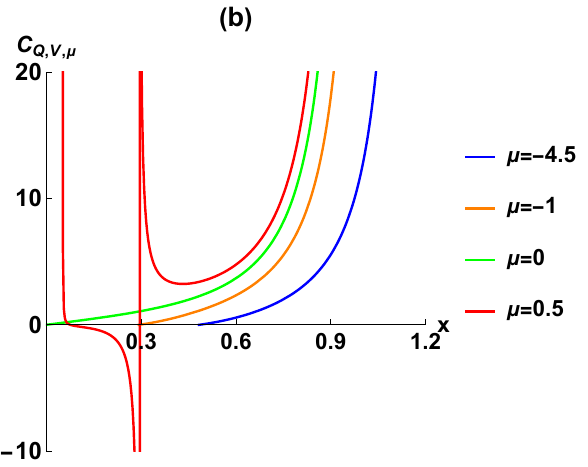}
 	\includegraphics[scale=0.55]{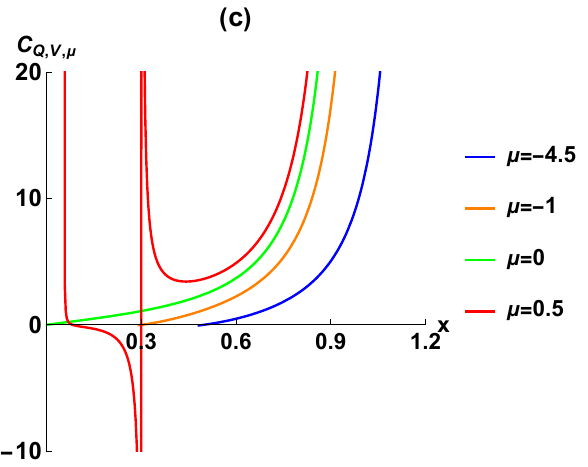}\hspace{0.5 cm}
 	\includegraphics[scale=0.55]{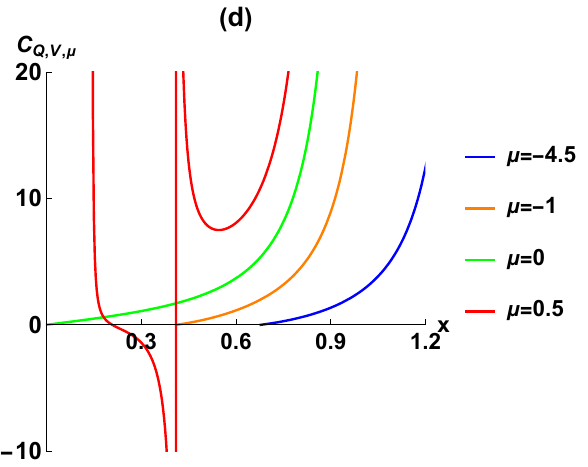}
 	\caption{Heat capacities, for fixed $(\tilde Q,{\cal V},\mu)$ ensemble. $R=0.1,\,\, \tilde{Q}=1$. The $\zeta$ of (a) to (b) corresponds to $0, 0.075, 0.3, 0.8$, respectively.}
 	\label{figure 7}
 \end{figure}

 The images in the bottom row of Figure \ref{figure 7} correspond to $(\tilde Q,{\cal V},\mu)$ ensemble. In this ensemble, the blue and orange curves represent $\mu < 0$, the green curve represents $\mu = 0$, the red curve represents $\mu > 0$. At $\mu \leq 0$, the heat capacity is always positive. At $\mu > 0$, the curve branches, with the smaller part of $x$ being the low-entropy phase and the larger part of $x$ being the high-entropy phase. The heat capacity of the high entropy phase is positive, and the CFT phase is always stable. When the curve representing the low-entropy phase is at a certain position, the heat capacity changes from positive to negative, and the CFT phase also changes from stable to unstable. Figure \ref{figure 3} corresponds exactly to the above situation. With the increase of global monopole, the $x$-value of low-entropy phase from stable state to unstable state increases. The range of low entropy phase and high entropy phase also increases with the increase of global monopole.
    
 \section{Discussion}
 \label{sec:discussion}
 We study the CFT thermodynamic phase behavior and critical phenomena of RN-AdS black hole with global monopole by means of AdS/CFT correspondence. In the second section, we introduce extended thermodynamics and CFT thermodynamics. In section 3, the thermodynamic phase transitions and critical phenomena of the three ensembles are discussed respectively. All of them lead to meaningful conclusions. Let's look at the results one by one.
 
 Firstly, the central charge $C$ and the chemical potential $\mu$ are introduced into the first law of thermodynamics as a new pair of conjugate thermodynamic quantities. In addition, changes in the cosmological constant and Newtonian constant are also considered. So that all the terms of the first law of thermodynamics correspond to the CFT. We obtain the first law of CFT thermodynamics through the holographic CFT dictionary, which corresponds to the thermodynamic quantities in the AdS background to the CFT background. 
 
 Secondly, we talked about fixing the ensemble $(\tilde Q,{\cal V},C)$, $(\tilde{\Phi},{\cal V},C)$ and $(\tilde Q,{\cal V},\mu)$. Critical phenomena appear in all three ensembles. In $(\tilde Q,{\cal V},C)$ ensemble, fixing the other parameters, a ``dovetail" image appears when the central charge is greater than the critical value of the central charge. After discovering this phenomenon, we only change the size of the charge, the other parameters remain unchanged, when the charge is less than the critical value of the charge, the ``dovetail" image also appears. In order to study the influence of global monopole on this phenomenon, we choose the cases of $C < C_{crit}$, and only change the size of global monopole. With the increase of global monopole, the size of the dovetail is basically unchanged(that is, the unstable region remain unchanged), but the position continues to move to the bottom left. In terms of the rate at which the unstable region increases. In the ensemble $(\tilde{\Phi},{\cal V},C)$, when $\tilde{\Phi} \geq \Phi_{c}$, the free energy is a smooth curve with no extreme value. At $\tilde{\Phi} < \Phi_{c}$, the curve has a maximum value. The curve is divided into two branches. The upper branch corresponds to the low entropy state, and the lower branch corresponds to the high entropy state. The branch corresponding to the high entropy state is truncated by the horizontal axis at some point and is represented as a first-order phase transition. Therefore, at $\tilde{\Phi} < \Phi_{c}$, we fixed the other parameters, increased the global monopole, and found that the point of the branch of the high entropy state truncated by the horizontal axis is moving to the left. The free energy also decreases with the increase of global monopole. It can be seen that the phase transition still exists in the presence of global monopole. We find that the Hawking temperature does not change with the global monopole. And $\phi_{c}$ doesn't change with the global monopole. In the ensemble $(\tilde Q,{\cal V},\mu)$, the fixed $\mu$ is the fixed thermal free energy per degree of freedom, and we chose some $\mu$ values. The ensemble then compares different states with different thermal free energies for each degree of freedom. When $\mu > 0$, the curve has two branches. The upper branch corresponds to a low entropy state, and the lower branch corresponds to a high entropy state. With the increase of global monopole, the temperature of the intersection between the upper branch and the horizontal axis decreases, and the value of the upper and lower branches at the same point also decreases. The range of high entropy state and low entropy state decreases with the increase of global monopole.

Thirdly, it can be found that there are both high entropy states and low entropy states in the above three ensembles. So we take $(\tilde Q,{\cal V},C)$ ensemble as an example to study the coexistence lines of high entropy state and low entropy state. We found that the slope of the coexistence line is always negative. We found that the slope of the coexistence line is always negative. When other parameters are fixed, the global monopole increases, and the slope of the coexistence curve increases. After that, we want to see how CFT thermodynamics differs from van der Waals fluids. Therefore, we draw the $p-{\cal V}$ picture of CFT thermodynamics. In the holographic CFT background, there is no critical phenomenon in $p-{\cal V}$ images. When Newton's constant is fixed, $p-{\cal V}$ criticality exists in the whole. When Newton's constant is allowed to vary, the $p-{\cal V}$ criticality disappears. This is not surprising because the slope of the coexistence curve of van der Waals fluids is positive. Therefore, the hot state of CFT with global monopole is still not equivalent to van der Waals fluid.

Finally, in order to correspond with the data in Chapter 1. We keep all parameters the same as in Figures \ref{figure 1}, \ref{figure 2}, and \ref{figure 3}. The heat capacity diagram of the corresponding ensemble is given. Therefore, the location of the phase transition, the transition between the low entropy state and the high entropy state, has been well reflected by the heat capacity diagram. The reflected results are consistent with the previous ones. Through the above discussion, we find that with the increase of global monopoles, the thermodynamic phase transition and critical phenomena of black holes are better reflected. At the same time, the AdS/CFT duality is like a key to open people's minds, so that people will have more and more better understanding of black holes.

\section*{Acknowledgments}
\noindent This work was supported by the Natural Science Foundation of Sichuan Province (Grant No. 2022NSFSC1833), and by Sichuan Science and Technology Program (Grant No. 2023ZYD0023), and by the Science Technology Department of Sichuan Province (Grant No. R21ZYZF0001).

%\bibliography{criticality}
\providecommand{\href}[2]{#2}\begingroup\raggedright\endgroup

\end{document}